\documentclass[10pt,amsmath,amssymb,aps,nofootinbib,notitlepage,prd,superscriptaddress]{revtex4-1}

\let\cite\citep

\usepackage{lineno}
\usepackage[T1]{fontenc}
\usepackage{aas_macros,color,graphicx,hyperref,mathrsfs,orcidlink}
\usepackage{fontspec}
\hypersetup{colorlinks=true,allcolors=blue}

\definecolor{fed_blue}{HTML}{07004D}
\definecolor{steel_blue}{HTML}{2D82B7}
\definecolor{steel_blue_dark}{HTML}{1C71A6}
\definecolor{aqua_marine}{HTML}{42E2B8}
\definecolor{dutch_white}{HTML}{F3DFBF}
\definecolor{light_coral}{HTML}{EB8A90}
\definecolor{light_coral_dark}{HTML}{BA5A60}

\hypersetup{
    colorlinks = true,
    linkcolor = purple,
    urlcolor  = steel_blue_dark,
    citecolor = steel_blue_dark,
    anchorcolor = black
}
\let\originalleft\left
\let\originalright\right
\renewcommand{\left}{\mathopen{}\mathclose\bgroup\originalleft}
\renewcommand{\right}{\aftergroup\egroup\originalright}

\begin{document}

\title{\textbf{A universal relation predicts the observed reversal of a supernova chirp}}

\author{Joseph R. Farah\,\orcidlink{0000-0003-4914-5625}}
\email{josephfarah@berkeley.edu}
\affiliation{Miller Institute for Basic Research in Science, 206B Stanley Hall, Berkeley, CA 94720, USA}
\affiliation{Department of Astronomy, University of California, Berkeley, CA 94720-3411, USA}

\author{Logan J. Prust\,\orcidlink{0000-0002-7174-8273}}
\affiliation{Center for Computational Astrophysics, Flatiron Institute, 162 Fifth Avenue, New York, NY 10010, USA}

\author{Hang Yu\,\orcidlink{0000-0003-4914-5625}}
\affiliation{eXtreme Gravity Institute, Department of Physics,
Montana State University, Bozeman, MT 59717, USA}

% \author{others}

\begin{abstract}
% \jrf{134 words, too long}
% Superluminous supernovae, among the brightest stellar explosions, are thought to be powered by newborn magnetars. One such event, SN 2024afav, "chirped", displaying sinusoidal, increasing-frequency ($P(99 \textrm{ d})\sim39\pm2$ d to $P(166\textrm{ d})\sim17\pm5$ d) light-curve modulations. The model explaining this chirp, a precessing disk falling onto the dimming magnetar, generalizes to a normalization-independent universal relation between the period derivative and the dimming rate of the light curve. This relation makes a striking new prediction: a period minimum at $t\approx135$ d post-explosion, fixed by the magnetar's spin-down, after which the chirp reverses. Testing this prediction with continued follow-up of SN 2024afav, we report the discovery of five new modulations in which the chirp reverses ($P(177 \textrm{ d})\sim24\pm4$ d to $P(208\textrm{ d})\sim38\pm2$ d) at $t=155\pm6$ d as the relation predicts. Applied to chirping events from upcoming surveys, the relation could identify the precession mechanism and disk physics of each engine.

Superluminous supernovae, among the brightest stellar explosions, are thought to be powered by newborn magnetars. One such event, SN 2024afav, ``chirped'', displaying sinusoidal, increasing-frequency ($P(99\textrm{ d})\sim39\pm8$ d to $P(166\textrm{ d})\sim17\pm5$ d) light-curve modulations. The model explaining this chirp, a precessing disk falling onto the dimming magnetar, generalizes to a normalization-independent universal relation between the period derivative and the dimming rate of the light curve. 
%This relation predicts a period minimum at $t\approx150$ d post-explosion, fixed by the magnetar's spin-down, after which the chirp reverses. 
Applied to the published chirp alone, with the spin-down time measured from the unmodulated light curve, the relation predicts that the chirp must halt near $t \approx 150$ d and reverse.
With continued follow-up of SN 2024afav, we report the discovery of five new modulations in which the chirp reverses ($P(177\textrm{ d})\sim24\pm4$ d to $P(208\textrm{ d})\sim38\pm2$ d) at $t=147\pm3$ d, as predicted. Applied to chirping events from upcoming surveys, the relation could identify the precession mechanism and disk physics of each engine. 
\end{abstract}

\maketitle

% pitch problem; SLSNe central engine, nature paper, key question (what does the system have to  obey independent of microphysics)
% Difficulty interpreting the chirp in this model due to degeneracies in physics, competing explanations, therefore we formalize into universal relation and examine only supported predictions
% FIGURE: all band light curve fit to magnetar model at pre-modulation epochs as well as dashed line showing fit of magnetar model to entire light curve (to demonstrate inability to explain modulations), reference methods for more details
The power source of hydrogen-poor (Type I) superluminous supernovae (SLSNe-I), which outshine their ordinary supernova counterparts by at least an order of magnitude, remains an open question \citep{Gal-Yam2017a,Quimby2019,Moriya2019}. A leading explanation is that they are powered by a central engine, such as a newborn rapidly spinning magnetar, which deposits its rotational energy into the expanding ejecta \cite{Kasen2010,Woosley2010}. Despite this model predicting a monotonic, smooth decline post-peak, a majority ($50-70\%$) of SLSNe-I show bumps in their light curves that the magnetar model cannot reproduce \cite{Lunnan2018,Hosseinzadeh2022,Chen2023}. These bumps have been alternately attributed to either engine variability \cite[e.g.,][]{Dong2023, Zhang2025} or collisions with circumstellar material (CSM) \cite[e.g.,][]{Lin2023}, with surveys unable to distinguish between the two competing mechanisms \cite{Chatzopoulos2019,Hosseinzadeh2022}. 

\begin{figure}
\centering
\includegraphics[width=0.5\textwidth]{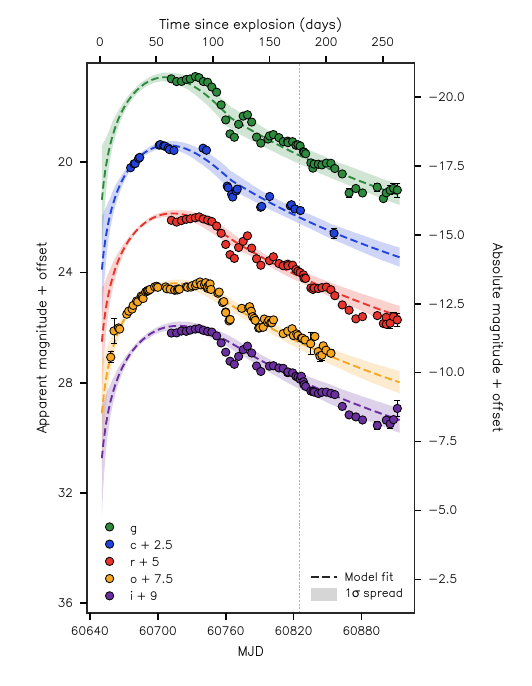}
\caption{\textbf{Magnetar model fit to SN 2024afav.} We fit the magnetar model of \cite{Kasen2010} to the full ATLAS and LCO light curve of SN 2024afav using \texttt{MOSFiT} \cite{Guillochon2018}. Uncertainties after MJD 60703 are inflated to absorb the impact of the modulations so they do not drive the fit. We find that the light curve out to $\sim250$ days post-explosion is well-described as powered by the spin-down of a magnetar with spin period $P_{\rm spin}\sim 4$ ms and magnetic field strength $B\sim0.5\times10^{14}$ G. However, the smooth post-peak decline of the magnetar model does not explain the presence of 6--7 post-peak modulations in the light curve. The vertical dashed line at $\approx175$ d post-explosion indicates the terminating epoch of the dataset of the previous analysis, which characterized the modulations at $t\lesssim170$ d as a ``chirp''. }
\label{fig:magnetar_vs_data}
\end{figure}

The discovery of SN 2024afav---the first ``chirping'' supernova, displaying in its light curve (\autoref{fig:magnetar_vs_data}) quasiperiodic modulations with an increasing frequency---presents a new trial for proposed mechanisms \cite{Farah2026Nature}. We attributed these modulations to a tilted fallback disk precessing around the magnetar, tying the bumps directly to the engine. However, interpreting the chirp in light of this model is challenging, due to the large number of degrees of freedom available. The neutron star's spin, magnetic field, and moment of inertia, as well as the structure of and accretion rate through the disk, all set the modulation period in combinations the light curve cannot disentangle. A good fit thus can only show that the engine picture is consistent with the data, not that the data require it. Similarly, a sufficiently structured CSM could theoretically reproduce any finite set of bumps \cite{Lin2023}. A stronger test of either hypothesis would therefore be a falsifiable prediction that can be made independently of the highly-degenerate degrees of freedom. Here, we reduce the engine model to the scalings that survive any choice of system parameters, derive a novel prediction from those alone, and test it against follow-up observations of SN 2024afav at late times. 

\section{A universal relation for the engine}
\label{sec:universal_relation_main}
% FIGURE: schematic view of the model
% relay framework assumptions, and ultimate relation; clarify what properties are no longer involved, reference methods for derivation
% explain universality of the linearity, how it can be broken, what it is resistant to
% conclude on hint at consequences (for next section)
Although the details of the engine configuration are challenging to measure individually, four requirements common to the engine picture are sufficient to uniquely determine how the chirp must follow the fading of the engine. First, the disk precesses at a rate which depends on a power law of inner-edge radius $r_i$ and the magnetar's angular momentum $J$, so that the precession period scales as $P\propto r_i^{K} J^{-q}$. The three physical mechanisms capable of producing this effect---Lense-Thirring, quadrupole-driven, and magnetic warp precession---all take this form, with differing exponents \cite{Mashhoon1984,Lai1999}. Second, the inner edge evolves with the location where the pressure of the magnetar wind, of luminosity $L_w$ (producing pressure $L_w/r_i^2$), balances the pressures of the disk \cite{Chashkina2019,Tamilan2025}, which we take in the most general case to be $\propto \dot{M}^{u} r_i^{-w}$ \cite{FKR}. Third, we assume that the accretion through the disk $\dot{M}$ is driven by fallback which declines as a power-law in time, $\dot{M}\propto t^{-\chi}$ \cite{Dexter2013} ($t=0$ assumed to be the explosion epoch, see \autoref{sec:universal_relation}). Finally, we assume the magnetar powers the wind extracting its rotational energy, with the rotational frequency $\Omega$ spinning down according to a braking index $n$, i.e., $\dot\Omega \propto \Omega^n$ \cite{Gunn1969}. We assume that the salient exponents stay constant during the chirp, and though we do not specify other parameters (e.g., disk $\alpha$) we assume they remain constant with time. Note that we do not require the assumption of a particular precession mechanism ($q$, $K$), disk pressure regime ($u$, $w$), braking mechanism ($n$), etc. 

Intriguingly, under this relatively broad set of assumptions, the new observables provided by the chirp are linked by a universal relation which contains none of the unknown system parameters (see \autoref{sec:universal_relation}). If we consider a measure of the ``chirping rate'', $C(t)\equiv - d\ln P/d\ln t$, and the diffusion-corrected ``dimming rate'' of the supernova, $D(t)\equiv - d\ln L_w / d\ln t$, the four assumptions above enforce,
\begin{align}
    C(t) = \frac{K\chi u}{w-2} - \left(\frac{K}{w-2} +\frac{q}{n+1}\right)D(t).
    \label{eq:CD}
\end{align}
Regardless of any particular choice of system parameters, the measured evolution of the supernova in the ($D, C$) plane must trace a straight line. This expected linearity is robust to most effects, including the choice of precession mechanism; the pressure regime of the disk; the spin-down mechanism, for any torque that scales as a power of the spin (magnetic dipole, magnetospheric, or gravitational-wave); a precession rate that also depends on the accretion rate, as in magnetic warping; and observational uncertainties (e.g., distance, explosion epoch, host extinction, etc.). The linearity demonstrated here is therefore a test of the proposed engine picture as a whole rather than of any particular realization of it. %It can be checked before the operating mechanisms are identified, so a curved or broken ($C, D$) track would invalidate the entire class of models, not merely one particular choice of physics. 
The same robustness extends beyond linearity, as any prediction that follows from Eq.~\ref{eq:CD} inherits its independence from the unknown parameters.

\section{The predicted period minimum}
\label{sec:main_predicted_period_minimum}
% major prediction: the period has a minimum which occurs at a particular time set by the spin-down time of the magnetar. (reference same methods section)
% Curvature is determined by spin-down time, so shape of chirp is effectively predicted by the pre-chirp light curve. 
% Extraordinary prediction: beyond t_min, chirp should reverse. 
Indeed, the consequences of \autoref{eq:CD} make a concrete prediction about chirping systems that require no particular choice of the unknown parameters. The linear construction of the relation directly implies that the chirp has a zero-crossing determined solely by the closure, at $D_{\rm min} = (w-2)^{-1}K\chi u / (K/(w-2) +\frac{q}{n+1})$, independent of any specific properties of the magnetar or the disk. In fact, the zero-crossing implies that the chirp halts and \textit{reverses} at the epoch $t_{\rm min}$ associated with $D_{\rm min}$, set by the magnetar spin-down. The physical origin of this reversal is the competing declines of the engine luminosity and the fallback accretion rate, whose interplay sets the disk radius (and thus the precession period), producing a turnaround when the magnetar engine shuts off after $\approx t_p$ (\autoref{sec:universal_relation}, \autoref{fig:reversal_physics}).

What makes this prediction remarkable is that the timing is largely driven by the spin-down time, which can be estimated from the light curve independently from the modulations. The system configuration ($K, w, \chi, \ldots$) sets the overall level (offset from $P=0$) and scale (magnitude between maximum and minimum period) of the chirp, but the shape (relative curvature at any two points; \autoref{sec:universal_relation}) of the periodicity evolution (and therefore the minimum) is essentially dictated by the smooth spin-down of the engine. An object observed to chirp past $t_{\rm min}$, or one that never executes a predicted reversal, falsifies the engine model outright. For SN 2024afav, whose published chirp ends just short of this regime, the prediction is unambiguous: continued follow-up should reveal a reversing chirp. 

The universal relation of \autoref{eq:CD} goes beyond predicting simply that the chirp will reverse, also forecasting characteristics such as the time of reversal and the spread of allowed period derivatives post-reversal (\autoref{sec:modulations_methods}). The prediction uses only two inputs, neither of which depends on the late-time modulations: (i) the five chirp periods previously reported in \cite{Farah2026Nature}, and (ii) the spin-down time measured from the smooth light curve. No extremum after $t \approx 170$ d is included in the fit. To produce a prediction independent of any particular closure, we fit \autoref{eq:CD} to the previously-reported chirp periods alone (determining the relation slope and intercept for the entire event without specifying the system configuration), with the spin-down time fixed to $t_p=87\pm 11$ d by the unmodulated light curve (\autoref{sec:magnetar_model_fit}, \autoref{fig:magnetar_corner}).  The constraint on the relation slope and intercept from the chirp-only periods, in conjunction with the spin-down time fit from the overall light curve, significantly limit the allowed behavior of the antichirp. The universal relation predicts that the chirp rate zero-crossing should occur at $t_{\rm min}\approx133$ d (rest frame, mode of $t_{\rm min}$ distribution from universal relation fit to chirp, see \autoref{fig:prediction}), before the period lengthens into an antichirp with an average period derivative of $\dot{P}=0.22^{+0.24}_{-0.11}$ between $t=157$ and $t=202$ d, with this interval chosen to match follow-up observations of SN 2024afav. This prediction is not simply an extrapolation of the chirp. While a simple linear fit to the chirp periods alone favors continued shortening (\autoref{fig:prediction}, dot-dashed line), the turnaround emerges organically once the chirp is fit with \autoref{eq:CD}, whose curvature is necessarily positive and whose minimum is timed by the spin-down. \autoref{eq:CD} thereby fixes three predictions in advance: that $\dot{P}$ changes sign, when it does so, and how steeply the period subsequently rises.

\section{Chirp reversal observed in SN 2024afav}
\label{sec:main_chirp_rev_obs}
% model reported in Nature paper (Sep 2025) has associated universal relation track that predicts period minimum at ~145 d with \pm in figure, completely inconsistent with continued linear-like decline. 
% followup observations continued, describe briefly (full description in materials and methods)
% additional modulations identified, covering 3-4 additional cycles; naive extrema extraction indicate lengthening period, power spectrum analysis indicates opposite behavior to t < 150d chirp. FIGURE: modulations identified on L(t) and highlighted in comparison to chirp/antichirp in residual plot and power spectrum. Multiple validations of chirp/antichirp in supplementary text
% Accounting for effects (e.g., diffusion timescale, etc., full description in materials and methods) and comparing to chirp-only prediction shows consistency with the predicted range. Proper model fit provides excellent match to period evolution and yields t_min highly consistnet with the pre-chirp median. FIGURE: chirp/antichirp comparison to prediction and model fit
The published chirp of SN 2024afav spans 7 extrema between about 50 and 170 days (times are from explosion in observer frame unless noted as rest frame) \cite{Farah2026Nature}. Over that interval, the period fell from $P=39\pm8$ d to $P=17\pm5$ d. The period evolution invites a simple linear extrapolation ($\dot{P}=-0.44\pm0.05$), and phenomenologically predicts that the period would keep shortening until it reaches zero near $t=180$ d. The universal relation forbids this, as did the model published with the chirp \cite{Farah2026Nature}, of which \autoref{eq:CD} is simply a generalization. In that model, the precession period decelerated, flattening toward $P\approx25$ d by $t=150$ d, rather than continuing to fall. Extending the model of \cite{Farah2026Nature} past the previously reported chirp predicts a clear period reversal (\autoref{fig:prediction}, panel A).

To test the prediction, we examine the data acquired during continued followup of SN 2024afav past the window analyzed for the published chirp. Through this follow-up, we obtained 109 further epochs in the $g$, $r$, and $i$ band with the Las Cumbres Observatory (LCO) global telescope network \cite{Brown2013} through MJD 60908 (262 days after explosion), alongside $c$ and $o$ photometry from the Asteroid Terrestrial-Impact Last Alert System (ATLAS; \cite{Tonry2018}). These epochs had been published earlier in \cite{Kumar2026} and were searched for modulations only after the precessing-disk model had been published \cite{Farah2026Nature}. We built the bolometric luminosity light curve following the same procedure as \cite{Kumar2026} (\autoref{sec:modulations_methods}). 

\begin{figure}
\centering
\includegraphics[width=\textwidth]{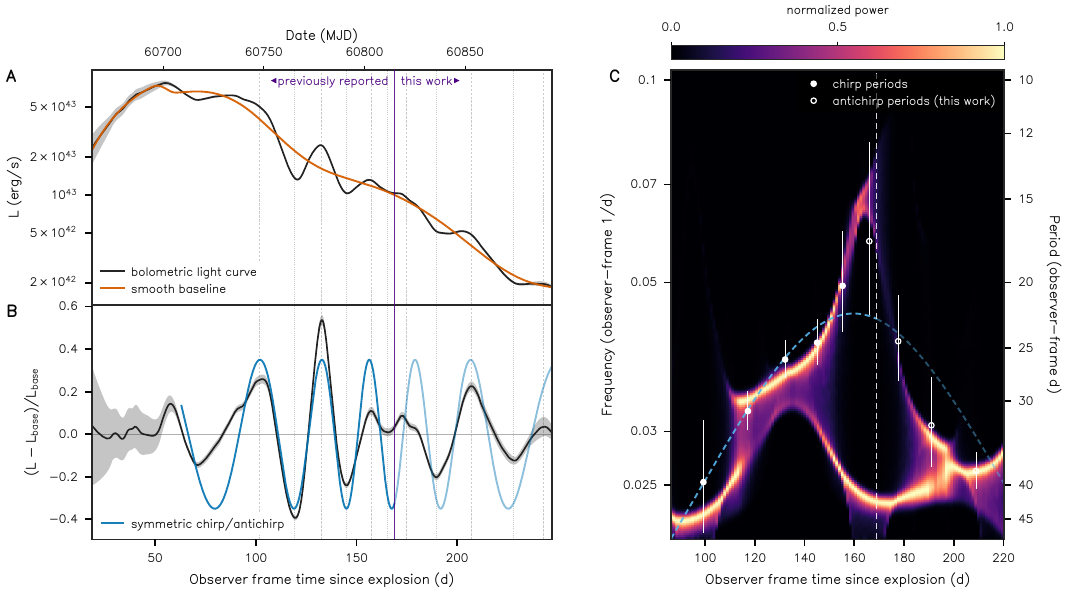}
\caption{\textbf{Timing of the modulations.} The post-peak light curve of SN 2024afav shows multiple significant modulations. Here, we characterize their timing using the bolometric luminosity as fit from the multi-band photometry following \cite{Kumar2026}. The chirp reported in the previous work (data left of the indicated purple line in panels A and B) is clearly recovered, with $\sim3$ cycles of a chirping modulation visible. After approximately $t=170$ d, several new modulations are visible in the residual (data subtracted from a smoothing cubic spline) and the unsubtracted light curve. The late-time modulations display extrema spaced increasingly far apart, reproduced in a weighted wavelet z-transform \cite{Foster1996} refined via the method of synchrosqueezing \cite{daubechies2011synchrosqueezed} (panel C, see \autoref{sec:validation_chirpantichirp_supp}). The evolution of the late-time modulations mirrors the originally-reported chirp, and its peaks and troughs are well-described by a sinusoid (blue curve, panel B; dashed blue line, panel C) whose period evolution is a symmetric parabola centered at $t\approx158$ d. We phenomenologically characterize this new set of modulations as an ``antichirp'' due to the lengthening period.}
\label{fig:timing_modulations_figure}
\end{figure}

\begin{figure}
\centering
\includegraphics[width=0.5\textwidth]{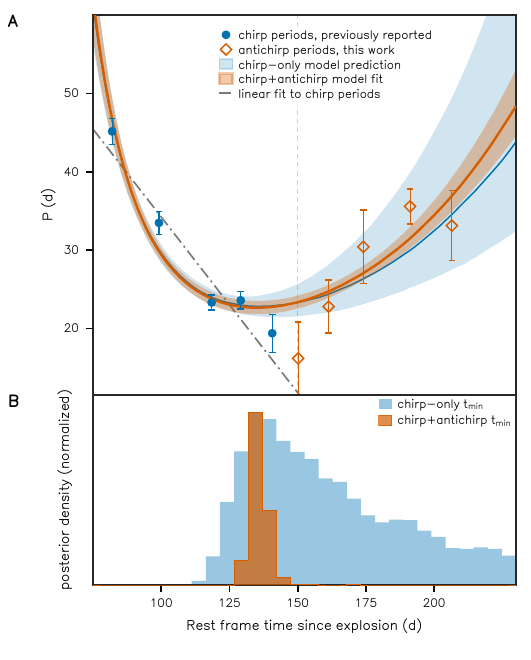}
\caption{\textbf{The predicted chirp reversal is observed.} We compare the predicted chirp reversal to the late-time modulations observed in SN 2024afav. The universal relation of \autoref{eq:CD}, when fit only to the chirp reported in previous work (panel A, blue points; end of previous analysis shown as dashed line), that a minimum should be present in the period at $t_{\rm min}\approx133$ d (rest frame; value inferred from mode of $t_{\rm min}$ posterior distribution inferred from the chirp only shown in panel B, blue) and should culminate in an antichirp with approximate period derivative $\dot{P}=0.22^{+0.24}_{-0.11}$ (panel A, blue line and shaded region). After correcting for diffusion through the ejecta and converting the observed extrema timing into the rest frame of SN 2024afav, we find that the new modulations are consistent with this prediction at $\lesssim1\sigma$ (panel B). The late-time modulations (orange points), when fit in conjunction with the chirp reported previously result in a minimum at $t_{\rm min}=132\pm3$ d (panel B, orange distribution) and produce $\dot{P}=0.28\pm0.12$ (orange line and spread). Finally, the period evolution displayed by the late-time modulations definitively rules out a continuation of the near-linear decline displayed by the periods which make up the chirp (dot-dashed line). Errorbars denote $1\sigma$ uncertainties on the period, constructed following \autoref{sec:modulations_methods}.}
\label{fig:prediction}
\end{figure}

The late light curve continues to show modulations. Relative to a cubic spline smoothing baseline (\autoref{sec:modulations_methods}), the residuals show five more extrema after $t=170$ d (\autoref{tab:extrema}), corresponding to roughly two and a half additional cycles (though we note that the modulations are also readily apparent in the non-subtracted light curve). The presence of a reversal is qualitatively visible in the spacings of the extrema (\autoref{fig:timing_modulations_figure}, panel B); before $t \approx160$ d, the extrema grow closer together, whereas after $t\approx 160$ d they drift apart. We identify identical modulations when constructing the residual using the (smooth post-peak) magnetar model (suggesting that the presence of the modulations is not sensitive to the flexibility in the smoothing spline baseline) and when using multiple methods of extrema identification. We further validate the behavior via a time-frequency decomposition (\autoref{fig:timing_modulations_figure} panel C; using the method of weighted wavelet z-transform, see \autoref{sec:validation_chirpantichirp_supp} for details). The dominant ridge in the power spectrum climbs from $\approx 0.025 \mathrm{\ d^{-1}}$ at $t\approx90$ d to $\approx 0.07 \mathrm{\ d^{-1}}$ at $t\approx170$ d, then clearly descends, tracing a reflection of the published chirp. The modulation frequency unambiguously rises and then falls---the chirp reverses. 

We tested the prediction on the antichirp periods extracted from the extrema of the modulations (\autoref{sec:modulations_methods}, \autoref{tab:periods}), correcting for the observer frame and for photon diffusion through the ejecta using the best-fit MOSFiT diffusion kernel. After the interval spanning the turnover ($16\pm 5$ d at $t=150$ d, rest frame), the period rises from $23\pm3$ d at $t=161$ d to $35\pm5$ d at $t=191$ d (rest frame). This behavior results in a slope of $0.28\pm0.12$, at the $\approx$48th percentile of the predicted distribution, and excluding the continued decline of the linear extrapolation at $\gg5\sigma$. Fitting the chirp and antichirp together with the model of \autoref{eq:CD} places the minimum at $t_{\rm min}=132\pm3$ d (rest frame), statistically consistent with the prediction. The behavior of the new modulations matches the prediction from the universal relation.

\section{Circumstellar interaction does not set the timing}
\label{sec:main_alternatives}
% primary competitive hypothesis is CSM, and indeed evidence of CSM interaction in spectra. 
% at this point would require extreme fine tuning of the CSM 
% we use distribution of SLSNe modulations to perform red noise analysis and find excluded to X% [in progress]; random bumps fail to reproduce chirp-antichirp behavior (full description in materials and methods)
% Additional point is now with universal relation, clear predictions can be made and scenarios ruled out; competing fine-tuned explanations can't achieve this yet (segue into further predictions in next section)

The primary alternative explanation for light curve undulations in SLSNe-I is interaction between the ejecta and CSM \cite{Liu2018,Hosseinzadeh2022,Lin2023}. Indeed, spectra of SN 2024afav show several features consistent with CSM interaction \cite{Kumar2026}, so CSM may well be present. However, the presence of a chirp-antichirp feature in the light curve invites the question of whether CSM is merely present or actively driving the observed modulations. We consider three features of the data that we argue indicate CSM is not the primary driver of the light curve modulations. 

First, the modulations are inconsistent with the stochastic variability seen in other SLSNe-I, also proposed to be driven by CSM interaction. We modeled the post-peak modulations of six SLSNe-I as red noise and generated $\sim5\times10^4$ realizations at the cadence and photometric uncertainties of SN 2024afav. Each realization was processed identically to the SN 2024afav dataset (see \autoref{sec:rednoise_methods} for additional details) and scored on two quantities: (i) the modulation power (measured via excess root-mean-square (rms) amplitude), and the chirp-antichirp parabola ``contrast'', a measure of the strength of the chirp-antichirp feature present. SN 2024afav is an outlier in comparison to the variability present in other SLSNe-I, with $\lesssim0.1\%$ of realizations displaying behavior as extreme as SN 2024afav in both quantities combined (\autoref{fig:rednoise_fig}). As bumps in similar SLSNe-I are commonly attributed to CSM interaction, this result implies that interaction-driven variability of the kind observed in SLSNe-I \cite{Hosseinzadeh2022} typically does not produce a chirp followed by an antichirp. Even when the red noise process is calibrated to SN 2024afav exclusively, realizations with comparable power and parabola contrast occur $\lesssim5\%$ of the time. 

Second, CSM shells placed at deliberate radii could potentially produce the extrema observed, but the physical scenario required to do so is contrived. A decelerating shock crossing shells should produce intervals that lengthen with time \cite{Chevalier1982} unless the shell spacing itself shrinks outward. Further, while the chirp requires spacings that contract outwards over several shells, the antichirp requires the reverse---spacing that expands (or stays constant while the shock significantly decelerates). Previous mechanisms proposed to explain possible regularity in CSM shell placement (e.g., pair instability, \cite{Lin2023}) do not easily explain the turnaround observed. Additionally, the data enforce that the transition must occur at a specific time, $t\sim150$ d. While this turnover time is predicted by our model, there is nothing in the progenitor's mass loss history that ties the spacing of the shells to this specific epoch. 

Finally, the shape of the modulations is inconsistent with previous models of CSM interaction. Forward and reverse shock collisions with shells produce highly asymmetric bumps with a rapid rise and slow decay \cite{Lin2023,Khatami2024}. The modulations of SN 2024afav---particularly during the chirp---are consistent with a sinusoid with a dynamic envelope \cite{Farah2026Nature}, which is inconsistent with the CSM interaction profile even before any attempt to match timing is performed. In the engine picture, by contrast, the period sequence, magnitude of the turnover, and reversal epoch all follow from \autoref{eq:CD} with no additional freedom. The proposed relation also makes testable predictions for events beyond SN 2024afav. 

\section{Discrete properties of the universal relation}
\label{sec:main5}
% exponents originate from discrete physics (K, disk regimes, n) so allowed lines form a comb.  (more discussion in supplementary information)
% C-D are essentially measured from the light curve (subtlety: requires undoing diffusion, which requires model fit. fitting spin-down time from early light curve and convertin D <-> t lets you do this without requiring the chirping portion of the dataset). 
% therefore by simply placing measured C-D on comb you can evaluate disk physics and precession mechanism without requiring a fit
% allows for falsifiability: deviations from linearity/curvature, or C-D falling outside of comb would indicate model is incorrect even if chirp appears present
% future sample of objects (e.g., simple Rubin rate) will fill in comb and indicate what disk physics/precession mechanisms are preferered; once these quantities are nailed down, more detailed measurements of the system (e.g., moment of inertia) may become possible.

A further property of the universal relation presented in \autoref{eq:CD} is that its coefficients are quantized, which provides a meaningful diagnostic for future events. The parameters of the system ($K$, $w$, $\chi$, $u$, $n$) are fixed by discrete physical choices (see \autoref{sec:discrete_supp}), as opposed to continuous tunable parameters. The precession mechanism sets $K$ and $q$; the spin-down of a magnetar with a dipole field sets $n=3$; the pressure and opacity regime of the disk sets ($u, w$); and the assumption of fallback sets $\chi=5/3$. The consequence of the quantization is that no physical parameter interpolates between the discrete values, and each combination as a result yields an isolated line, forming a ``comb'' in the $D$--$C$ plane. The line constructed by measuring $C$ and $D$ from the light curve can therefore identify the active precession mechanism and disk regime, and an observed track that falls between or beyond the lines effectively excludes the entire menu. 

\begin{figure}
\centering
\includegraphics[width=0.5\textwidth]{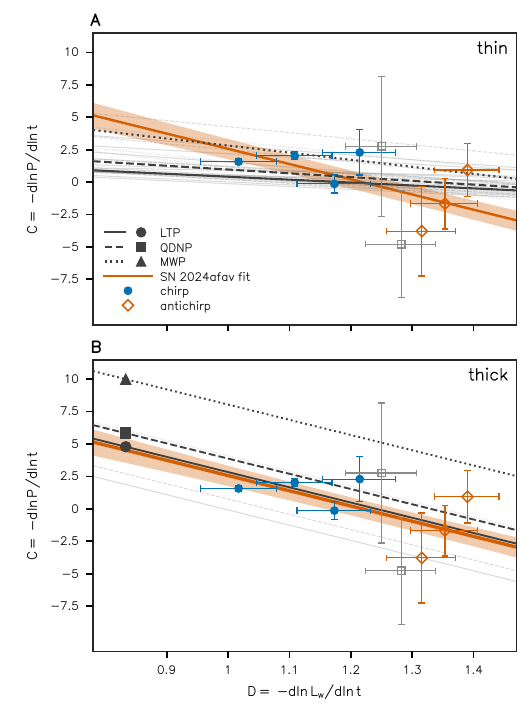}
\caption{\textbf{Discreteness of the universal relation.} The universal relation is constructed almost entirely from discretized quantities whose values represent specific configurations of the system. This quantization results in a finite set of lines against which the best-fit line traced by a supernova in the ($D$, $C$) space can be checked without requiring a fit, providing both a diagnostic as well as a falsifiability criterion. Indeed, we find that no line on the thin-disk (radiatively-cooling) comb (panel A) produces a slope as steep as SN 2024afav, statistically rejecting the thin-disk prescription for this object (best-fit closure---quadrupole-driven nodal precession (QDNP)---gives reduced $\chi^2=9.8$, $p=3\times10^{-15}$). By contrast, a thick-disk (advective) prescription (panel B) can reproduce the observed slope and intercept in SN 2024afav (at $\lesssim0.2\sigma$) with Lense-Thirring precession (LTP). Future chirping objects can be classified and characterized in this way. In Panel A, black lines indicate the best-match closure for each mechanism, while gray lines indicate the remaining closure menu (see \autoref{tab:thin_menu}). In Panel B, black lines indicate the $\chi=5/3$ \cite{Dexter2013} configurations, while gray lines indicate their $\chi=4/3$ \cite{Metzger2008} counterparts. } 
\label{fig:ur_comb_thick_thin_comp}
\end{figure}

Our application of this method to SN 2024afav demonstrates the discriminating power of the discreteness test (\autoref{fig:ur_comb_thick_thin_comp}). The chirp alone lies within the allowable range spanned by thin-disk lines, consistent with the reported \cite{Farah2026Nature} ability of the published thin-disk model to reproduce it. However, the chirp spans too narrow a range in $D(t)$ to effectively constrain the slope without the reversal (\autoref{fig:ur_comb_thick_thin_comp}). The chirp and antichirp combined constrain the slope to $dC/dD=11.6\pm1.6$, significantly steeper than any line on the thin-disk comb. As a result, no thin-disk closure is consistent with the measured relation, the best-fitting closure (QDNP/gas–Kramers/gas) gives reduced $\chi^2 = 9.8$ ($p = 3\times10^{-15}$). The track is linear (curvature consistent with zero at $\lesssim1\sigma$), consistent with the relation, but the inconsistency with the comb suggests that no radiatively-cooled thin disk can reverse the chirp as sharply as observed.  

Estimates of the fallback rate at late times exceed the neutron star’s Eddington rate by orders of magnitude \cite{Chevalier1989_accretion,FKR}, even in the thin-disk fit to the chirp alone \cite{Farah2026Nature}. As a result of the high accretion rate, the inner disk should be advective, cooling inefficiently and losing mass to outflows inside the photon-trapping radius \cite{Begelman1979,Blandford1999}. Adopting the self-similar structure of such a flow, with the inflow rate declining inward as $\propto r^p$ and the trapping radius scaling with the fallback supply, the disk pressure scales as $\dot{M}^{1-p} r^{p-5/2}$, giving $u = 1 - p$ and $w = 5/2 - p$ with $p$ continuous (though a chirp-antichirp requires $p < 1/2$ by construction). Remarkably, the universal relation retains its discriminating power despite the presence of a continuous parameter. Because the same $p$ sets both the slope and the intercept, every line belonging to a given precession mechanism passes through one fixed point in the $(D, C)$ plane (\autoref{sec:discrete_supp}), at $D = \chi/2$ and $C = K\chi - q\chi/[2(n + 1)]$. This point is set by the precession mechanism, fallback index, and braking index; the outflow strength only rotates the line about the pivot point. Discreteness survives as a pivot point for each mechanism instead of a comb of lines, as the measured line must pass through one of them. We find that SN 2024afav passes almost directly through the Lense–Thirring point (offset $\lesssim0.2\sigma$), is inconsistent with the quadrupole point ($\gtrsim1\sigma$), and rules out magnetic-warping (by $\gtrsim5\sigma$). The slope as a consequence then reports the outflow index, $p \sim 0.46$ for Lense–Thirring precession, at the weak-outflow end of the values found in simulations \cite{Yuan2012}. This test is selective for different model choices; for example, had the outflow zone ended at a fixed radius rather than at the trapping radius, no mechanism would reproduce the modulations (\autoref{fig:ab_vs_b_supp}). 

\section{Summary of the observed chirp reversal}

From the universal relation and continued follow-up of SN 2024afav, we conclude that the chirp reverses where the engine model predicts. The relation holds for any precession mechanism, disk pressure regime, or braking index, and requires the chirp to halt at a time set by the spin-down of the magnetar. Fit to the published chirp alone, the new relation predicts a period minimum at $133 \pm 20$ d (rest frame). The late-time light curve shows five new extrema with a period minimum at $132 \pm 3$ d (rest frame), after which the period lengthens from $23 \pm 3$ d to $35 \pm 2$ d in a clear antichirp. SLSN red noise cannot reproduce this behavior to any reasonable probability ($\lesssim0.1\%$ of simulated realizations). The sharpness of the reversal (relation slope $dC/dD = 11.6 \pm 1.6$) excludes every radiatively-cooled thin disk and points to a hot, advective inner flow. Applied to future chirping events, the same test requires no model fit, as each event will either land on the thin-disk menu, identifying its precession mechanism directly, or, like SN 2024afav, constrain the structure of its disk. We conclude that the unprecedented chirp and antichirp of SN 2024afav are driven by the magnetar engine, whose output is modulated by a precessing disk supported by the magnetar’s wind.

\bibliography{ref}

\begin{acknowledgments}

\noindent We gratefully acknowledge D. A. Howell for acting as an internal referee for this paper. We thank D. Kasen, W. Lu, R. Margutti, D. Gates, R. Chornock, J. Delgado, R. Narayan, and D. Pesce for helpful discussions and feedback. We thank M. Haller, E. Abate, and A. Mei for reviewing the manuscript and providing comments. This work makes use of data from the Las Cumbres Observatory (LCO) global telescope network and the Asteroid Terrestrial-impact Last Alert System (ATLAS) project, both of which make use of instruments on Haleakala. We respectfully acknowledge the profound cultural significance and enduring reverence of the summit of Haleakala to the indigenous Hawaiian community and we are grateful for the opportunity to study the heavens from this mountain. 
\vspace*{0.3cm}

\noindent \textbf{Funding:} J.R.F. is grateful to the Miller Institute for Basic Research in Science at UC Berkeley for supporting this work. L.J.P. acknowledges support from The Flatiron Institute, which is supported by the Simons Foundation. H.Y. is supported by NSF through Awards PHY-2308415, PHY-2541579 (CAREER Award), and AST-2606125, and Montana NASA EPSCoR Research Infrastructure Development under award No. 80NSSC22M0042. The LCO group is supported by NSF grants AST-1911225 and AST-1911151. The ATLAS project is primarily financed to search for near-Earth asteroids through NASA grants NN12AR55G, 80NSSC18K0284 and 80NSSC18K1575; by-products of the near-Earth object search include images and catalogues from the survey area. The ATLAS science products have been made possible through the contributions of the University of Hawaii Institute for Astronomy, the Queen’s University Belfast, the Space Telescope Science Institute, the South African Astronomical Observatory and the Millennium Institute of Astrophysics (MAS), Chile.
\vspace*{0.3cm}

\noindent \textbf{Author contributions:} (following CRediT system) \textit{Conceptualization}: J.R.F. \textit{Data curation}: J.R.F. \textit{Formal analysis}: J.R.F., L.J.P. \textit{Investigation}: J.R.F. \textit{Methodology}: J.R.F., L.J.P., H.Y. \textit{Project administration}: J.R.F. \textit{Software}: J.R.F., L.J.P. \textit{Validation}: L.J.P., H.Y. \textit{Visualization}: J.R.F., L.J.P. \textit{Writing (original draft)}: J.R.F. \textit{Writing (review \& editing)}: J.R.F., L.J.P., H.Y.
\vspace*{0.3cm}

\noindent \textbf{Competing interests:} There are no competing interests to declare.
\vspace*{0.3cm}

\noindent \textbf{Data, code and materials availability:} The data used in this analysis are made publicly available on WiseREP. Code used to run the analysis is available on GitHub. 
\end{acknowledgments}

\clearpage
\appendix

\section{Materials and methods}
\label{sec:methods}

\subsection{Observations of SN 2024afav}
\label{sec:observations_methods}
% brief description of original observations (reference nature paper) with more detailed description of follo-wup observations
SN 2024afav (RA 12h49m12.050s, Dec-18$^\circ$06'12.61'', J2000) was discovered by L-GOTO on 2024 December 12.7 UTC \cite{Kumar2024} and classified as a SLSN by ePESSTO+ in January 2025 \cite{deWet2025}. We adopt $z=0.072$, measured from the narrow host lines. Assuming flat cosmology ($\Omega_M=0.286$, $H_0=69.6 \mathrm{\ km \ s^{-1} \ Mpc^{-1}}$), we adopt a distance to SN 2024afav of 327 Mpc. Following \cite{Farah2026Nature,Kumar2026}, we correct all photometry for Milky Way extinction with $E(B-V)_{\rm MW}=0.0527\pm0.0029$ mag. As in \cite{Farah2026Nature}, we apply no host extinction correction, as the host offset combined with an upper limit of $E(B-V)H \lesssim0.04$ mag suggest the host contribution is negligible. 

The additional (i.e., beyond the chirp reported in \cite{Farah2026Nature}) photometry analyzed in this paper has already been reported in \cite{Kumar2026} and was obtained via ATLAS and LCO. ATLAS observed the field on alternate nights in $c$-band ($\approx420-650$ nm) from before discovery through MJD 60853, and roughly once per week in $o$-band ($\approx560-820$ nm). We obtained forced photometry \cite{Tonry2018} on ATLAS difference images in both bands. Multiple exposures from the same night were combined to increase signal-to-noise on the detections for each epoch. ATLAS photometric data points were calibrated to the Pan-STARRS catalog \cite{Chambers18}.

LCO monitoring began near the middle of the ``plateau'' present in the early light curve, on MJD 60711, as part of follow-up with the Global Supernova Project. We used the 1m and 2m network \cite{Brown2013} in $g$-, $r$-, and $i$-band on an approximately daily cadence. Images were reduced with the \texttt{lcogtsnpipe} pipeline and calibrated to the Sloan Digital Sky Survey catalog AB magnitudes.

\subsection{Fit to the magnetar model}
\label{sec:magnetar_model_fit}
% fits of the overall magnetar model but particularly to the early light curve 
% FIGURES: corner plots for both fits
The peak luminosity of SN 2024afav ($M\approx-20.5$) and the $\sim50$ day rise time are typical of magnetar-powered SLSNe-I. We fit the magnetar model of \cite{Kasen2010} using the Modular Open-Source Fitter for Transients (MOSFiT; \cite{Guillochon2018}) to all 811 ATLAS ($c$ and $o$) and LCO ($g$, $r$, and $i$) photometric measurements acquired during observations of the target. As the model declines smoothly post-peak and cannot reproduce modulations of any kind, we use this fit to estimate engine parameters independently of the chirp-antichirp and to demonstrate that the modulations are not a property of the smooth central engine. To perform the fit, we vary the explosion epoch, spin period, dipole field, ejecta mass, ejecta velocity, gamma-ray opacity, and photospheric temperature floor, fixing all remaining parameters (e.g., neutron star mass, optical opacity, etc.) to reasonable values. 

The magnetar model is sensitive to post-peak behavior, but the light curve is significantly impacted by the presence of the modulations at $t \gtrsim 50$ d. The modulations are a correlated astrophysical signal, not measurement noise, and thus we wish to prevent them from driving the fit. To achieve this, we add a fixed uncertainty of $\sigma_{\rm mod}=0.515$ (equal to $\approx1/2$ the peak-to-trough amplitude of the largest modulation) in quadrature to the photometric uncertainties associated with measurements after the first appearance of the modulations (approximately MJD 60703). We additionally incorporate an intrinsic scatter term $\sigma_{\rm eff}$ which will be dominated by $\sigma_{\rm eff} \ll \sigma_{\rm mod}$ at $t > $MJD 60703 but impacts the reliability of the fit on the smooth rise. The favored intrinsic scatter term favors $\sigma_{\rm eff}=0.059\pm0.008$ mag, a largely negligible scatter. We find this approach is ultimately conservative; for the median model, $\chi^2=0.43$ for all 811 points ($\chi^2=1.14$ for pre-modulation points whose uncertainties are dominated by $\sigma_{\rm eff}$, and $\chi^2=0.29$ for modulation-impacted points dominated by $\sigma_{\rm mod}$). 

\begin{figure}
\centering
\includegraphics[width=\textwidth]{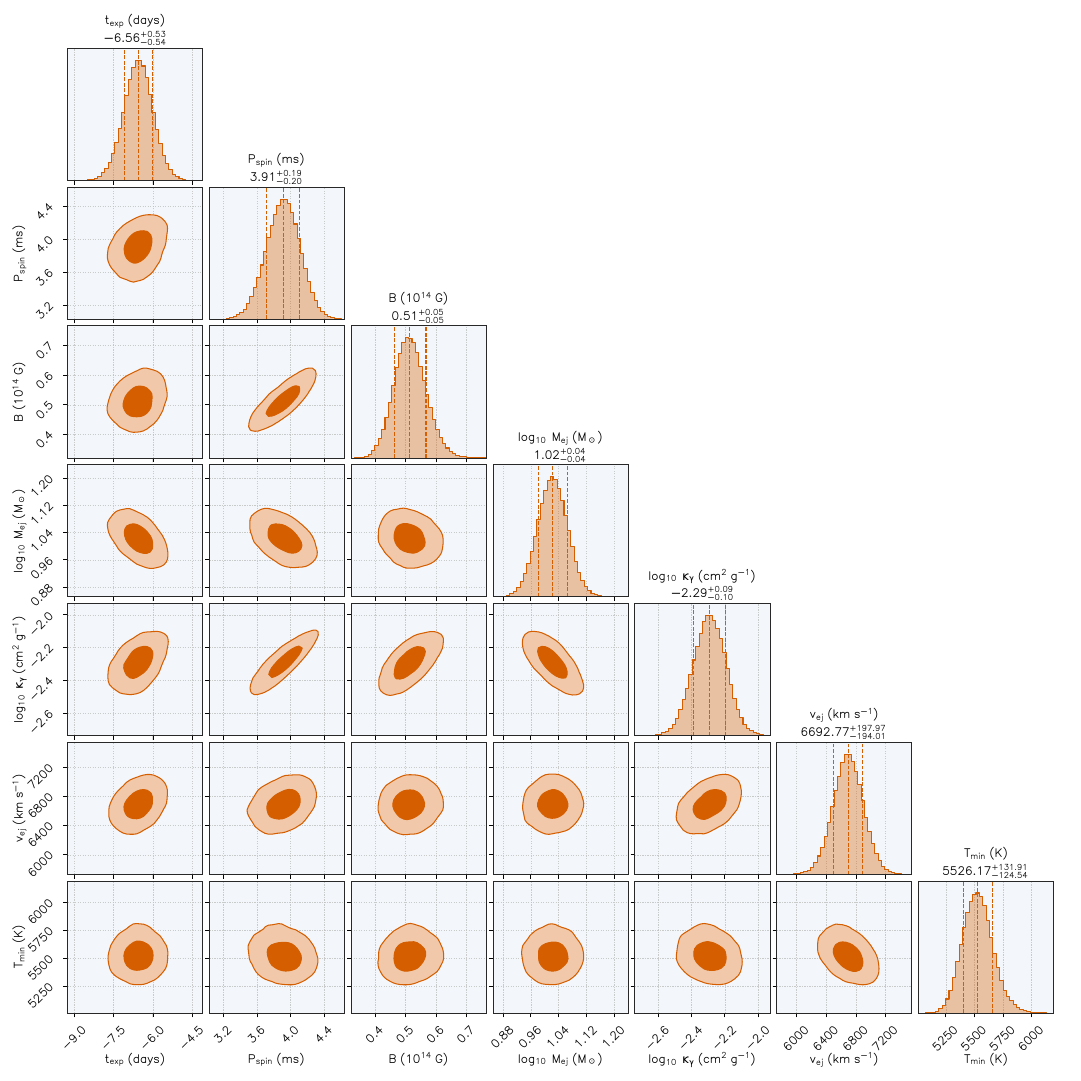}
\caption{\textbf{Magnetar model fit posterior.} We show the results of the magnetar model fit described in \autoref{sec:magnetar_model_fit} via a corner plot of the posterior conditioned on the full SN 2024afav light curve. }
\label{fig:magnetar_corner}
\end{figure}

Having adjusted the uncertainties to minimize the impact of the modulations, we condition the magnetar model on the photometric dataset. We use Markov-Chain Monte Carlo (MCMC) to sample the posterior with 72 walkers for $\approx10^4$ steps and discard the first 40\% as burn-in. Convergence is assessed following \cite{Farah2025a}; the split-chain Gelman-Rubin statistic is $\leq1.03$ for every parameter. We quote medians of the posterior distributions for each parameter with uncertainties drawn from the 16th-84th percentile intervals. This simple approach is justified as the posterior is well-behaved, with fast convergence and relatively symmetric posteriors. The results are shown in \autoref{fig:magnetar_vs_data} and the corresponding posterior is shown in \autoref{fig:magnetar_corner}. We find that the fit can reproduce the rise, peak, and mean decline in all five bands with $P_{\rm spin}=3.9\pm0.2$ ms, consistent with the previously reported fit up to $1\sigma$; and magnetic field strength $B=(0.51\pm0.05)\times10^{14}$ G, about a factor of $\approx2$ smaller than previously reported. The ejecta mass and ejecta velocity are similar to previously reported results as well. The fit favors an explosion epoch of MJD $60646.1\pm0.5$, which is $50.3\pm0.5$ rest-frame days prior to maximum light. We adopt this explosion epoch as the reference epoch for all light curves in this analysis. The best-fit parameters imply a rotational energy of $1.9\times10^{51}$~erg and a spin-down time (as input into the period law of \autoref{eq:CD}) of $t_p = 87\pm11$ d.

\subsection{Construction of the universal relation}
\label{sec:universal_relation}
% full derivation of universal relation and notes on subtleties 
We consider a disk (any regime) suspended on the wind of a central magnetar, and assume that the precession period scales with the radius of the disk's inner edge and the angular momentum of the central engine,
\begin{equation}
  P = \mathcal{A}\, r_i^{K} J^{-q},
  \label{eq:h1}
\end{equation}
with $\mathcal{A}$, $K$, and $q$ constant over the window. This form covers a broad range of precession mechanisms, including quadrupole-driven nodal precession \cite{Katz1996}, magnetic warp precession (MWP; \cite{Lai1999}), and Lense--Thirring precession \cite{Iorio2024book,Mashhoon1984}, for which $K>0$ and $q\geq0$; we neglect spin-up of the magnetar. We further assume that the inner edge evolves with the balance point between the magnetar wind $L_w$ and the combined disk pressures, which regardless of regime are power laws in radius and accretion rate~\cite{FKR},
\begin{equation}
  \frac{L_w}{r_i^2} = \mathcal{C}\,\dot M^{u} r_i^{-w},
  \label{eq:h2}
\end{equation}
with $u$ and $w$ constant (their non-constancy has a direct observable consequence, discussed below). The accretion rate follows fallback, $\dot M=\dot M_0 (t/t_0)^{-\chi}$, with $\chi=5/3$ at late times~\cite{Dexter2013}; although $\chi$ and $u$ enter only as a product, they are drawn from different menus (fallback physics and the disk response, respectively). Finally, the magnetar spins down with braking index $n$, $\dot\Omega=-\mathcal{K}\Omega^{n}$~\cite{Gunn1969,Ostriker1969}, driving a wind $L_w=-I\Omega\dot\Omega$ with $J=I\Omega$~\cite{Kasen2010}. Times $t$ are measured from magnetar birth ($\approx$ explosion), and the observables are $C\equiv-d\ln P/d\ln t$ for the engine-frame period and $D\equiv-d\ln L_w/d\ln t$ for the wind luminosity prior to diffusion. We define $s\equiv1/(w-2)$, $A_{\rm UR}\equiv\chi K s u$, and $B_{\rm UR}\equiv Ks+q/(n+1)$.

Rearranging Eq.~\ref{eq:h2} (for $w\neq2$) isolates $r_i=(\mathcal{C}\dot M^u/L_w)^{s}$. Integrating the spin-down law for $n>1$ gives $\Omega^{1-n}=\Omega_0^{1-n}(1+\tau)$, where $\tau\equiv t/t_p$ and $t_p\equiv[(n-1)\mathcal{K}\Omega_0^{n-1}]^{-1}$ is the doubling time of $\Omega^{1-n}$, so that $\Omega=\Omega_0(1+\tau)^{-1/(n-1)}$ and
\begin{equation}
  L_w=L_0(1+\tau)^{-\frac{n+1}{n-1}},\qquad
  D=\frac{n+1}{n-1}\frac{\tau}{1+\tau},
  \label{eq:D}
\end{equation}
with $L_0\equiv I\mathcal{K}\Omega_0^{n+1}$. $D$ is simply a function of the braking index and spin-down time, both of which we expect to be constant, along with $t$, and so serves as an almost specifics-independent clock of the system. Likewise $d\ln J/d\ln t=-D/(n+1)$, which holds simply because $L_w\propto\Omega^{n+1}$ and $J\propto\Omega$, so long as $I$ and $\mathcal{K}$ are constant over the chirping window. Substituting the fallback rate and Eq.~\ref{eq:D} into $r_i$, and $r_i$ and $J=J_0(1+\tau)^{-1/(n-1)}$ into Eq.~\ref{eq:h1}, we obtain
\begin{equation}
  P(t)=P_{\rm ref}\,(t/t_0)^{-A_{\rm UR}}(1+\tau)^{(n+1)B_{\rm UR}/(n-1)},\qquad
  P_{\rm ref}\equiv\mathcal{A}\,\mathcal{C}^{Ks}\dot M_0^{Kus}L_0^{-Ks}J_0^{-q},
  \label{eq:poft}
\end{equation}
and therefore
\begin{equation}
  C = A_{\rm UR} - B_{\rm UR}D = \chi K s u - \left(Ks+\frac{q}{n+1}\right)D.
  % \label{eq:CD}
\end{equation}
The relation is linear, and the line communicates only the exponents setting the disk, precession, and magnetar physics; it is independent of the specific values of the system. Explicitly, none of the clock constant $\mathcal{A}$ ($G$, $c$, geometric factors), the disk constant $\mathcal{C}$ ($\alpha$, opacity, mean molecular weight, the fraction of the wind intercepted by the rim), the fallback normalization $\dot M_0$, the engine constants $\mathcal{K}$ and $\Omega_0$ (equivalently $B$-field, birth period, and radius), the moment of inertia, the tilt, or any constant multiplicity of bumps per precession cycle ($P\to P/N$ rescales only $P_{\rm ref}$) enters $A_{\rm UR}$ or $B_{\rm UR}$. A fixed ratio of outer to inner disk radius $r_o/r_i$ is absorbed into $\mathcal{A}$, and if $r_o/r_i\gg1$ with $r_o$ tracking a power of $r_i$, it is absorbed into an effective $K$. The linearity therefore serves as a universal and meaningfully-falsifiable statement about supernovae powered by engine-clocked disk systems.

Deviations from linearity arise trivially if any exponent ($K$, $s$, $u$, $\chi$, or $q$) evolves over the chirp window. An offset $\delta$ in the explosion epoch (typically $\pm0.5$\,d) gives $t'=t-\delta$ and $d\ln t/d\ln t'=1-\delta/t$, so that $C'=(1-\delta/t)C$, $D'=(1-\delta/t)D$, and $C'=(1-\delta/t)A_{\rm UR}-B_{\rm UR}D'$: the slope is unchanged and the intercept is rescaled, with the time dependence of the rescaling introducing curvature only at $\mathcal{O}(\delta/t)\lesssim1\%$ for $t\sim10^2$\,d. %% original: linearity exactly preserved
Precession mechanisms whose normalization depends on the disk surface density, e.g.\ MWP with $\Omega_{\rm prec}\propto\Sigma^{-1}$ and $\Sigma\propto\dot M^{-m}$~\cite{Lai1999}, generalize Eq.~\ref{eq:h1} to $P\propto r_i^K\dot M^{-m}J^{-q}$, which yields $C=\chi(Ksu-m)-B_{\rm UR}D$ and leaves the linearity unchanged; because MWP has no intrinsic dependence on the magnitude of the spin vector, however, such a chirp would not probe $J$.

The framework further implies a falsifiable period minimum. Setting $C=0$ gives $D_{\rm min}=A_{\rm UR}/B_{\rm UR}$, and with $\Phi\equiv\tau/(1+\tau)$ and Eq.~\ref{eq:D},
\begin{equation}
  \Phi_{\rm min}=\frac{A_{\rm UR}(n-1)}{B_{\rm UR}(n+1)},\qquad
  \tau_{\rm min}=\frac{A_{\rm UR}(n-1)}{B_{\rm UR}(n+1)-A_{\rm UR}(n-1)}.
  \label{eq:tau_min}
\end{equation}
Using $d\Phi/d\ln t=\tau/(1+\tau)^2$, the curvature
\begin{equation}
  \frac{d^2\ln P}{d(\ln t)^2}=-\frac{dC}{d\ln t}=\frac{n+1}{n-1}B_{\rm UR}\,\Phi(1-\Phi)
  \label{eq:curv}
\end{equation}
is necessarily positive for $n>1$, $B_{\rm UR}>0$ (i.e.\ $Ks>-q/(n+1)$), and $\tau>0$, so the extremum is a minimum. %% original: n != 1; n<1 flips the sign
A chirp observed with a period continuing to decrease beyond $\tau_{\rm min}$ therefore falsifies the model for that system. In the edge case $B_{\rm UR}(n+1)<A_{\rm UR}(n-1)$, $\tau_{\rm min}<0$, there is no minimum at $t>0$, and the period declines monotonically; for $A_{\rm UR}<0$, $\tau_{\rm min}\in(-1,0)$ and the period rises monotonically. %% A<0 branch added
To apply the test, $n$ and $t_p$ are taken from a magnetar fit to the light curve ignoring the modulations (following the uncertainty-inflation method in \autoref{sec:magnetar_model_fit}; a joint magnetar+chirp fit is circular by construction), and $\ln P=-A_{\rm UR}\ln t+\frac{n+1}{n-1}B_{\rm UR}\ln(1+t/t_p)+{\rm const}$ is fit directly to the periods, with the full covariance of the correlated $A_{\rm UR}$ and $B_{\rm UR}$ propagated into $\tau_{\rm min}$. If the period is observed shortening up to some last epoch $t_{\rm last}$, the model requires $t_{\rm min}>t_{\rm last}$; a posterior placing $t_{\rm min}<t_{\rm last}$ at significance is direct falsification.

The predictive power of this test comes from fixing $t_p$ using data unrelated to the chirp. By Eq.~\ref{eq:curv}, the ratio of the curvature at any two epochs cancels $\frac{n+1}{n-1}B_{\rm UR}$ and depends only on $\Phi$, itself only a function of $t_p$; the model thus states how the chirp's deceleration must change across the window before any period is fitted, while $A_{\rm UR}$ and $B_{\rm UR}$ set only the level and scale and a single pair must serve the entire window. If the observed chirp decelerates much harder early than the curvature allows (the period ``collapses'' and then ``coasts''), the fit undershoots $C(t)$ and reports a minimum-period epoch while the period is still decreasing: an inconsistency between chirp and light curve that falsifies the model even when the fit is satisfactory by eye (i.e.\ with a good $\chi^2$). Conversely, since a poor $\chi^2$ does not distinguish an incorrect engine from a correct engine described by an oversimplified model, the minimum-epoch check provides a parameter-light self-consistency assessment beyond goodness-of-fit.

\subsection{Assessment of modulations}
\label{sec:modulations_methods}
% Full description of construction of periods and associated uncertainties, and how they are de-diffused, SNR weighted, etc. 
% POSSIBLE FIGURE: luminosity + per-filter modulations with bands highlighting uncertainties, or derivative to identify modulations?
We characterize the modulations in the bolometric light curve. We reproduce the blackbody bolometric luminosity evolution reported in \cite{Kumar2026} from the LCO ($g, r, i$) and ATLAS ($o, c$) bands with the open-source package \texttt{extrabol} \cite{extrabol}. In this section, we report times in observer-frame days since explosion (MJD 60646, from the magnetar fit, \autoref{sec:magnetar_model_fit}) and do not yet correct for diffusion or the rest-frame. 

First, we separate the modulations from the overall evolution of the supernova by fitting a smooth baseline to the light curve. We choose a cubic regression spline with five interior knots spaced 34 d apart following \cite{Kumar2026}. The choice of knot spacing can impact the modulation recovery, as a more flexible baseline could absorb modulations, while a less flexible baseline may not properly detrend the data. We vary the number of knots and their spacing around the identified optimal choice and find that the residual changes by on average $\lesssim0.04$ and at most $\sim0.08$, well below the modulation amplitudes ($\sim0.1-0.5$ mag). 

We identified 12 extrema (six peaks and six troughs) after $t=63$ d in the residual. The epoch associated with the extremum is first located by eye and then refined by a parabola fitted to the nearby modulation ($\approx35\%$ of the local half-cycle on each side). The modulations recovered from the bolometric luminosity using this procedure agree with the previously-reported chirp modulations (which were extracted directly from the per-filter photometric data) with a mean absolute offset of $\sim0.3$ d. In addition to the uncertainty from the quadratic fit, we add systematic uncertainties in quadrature based on (i) scatter in extraction using our pipeline as a result of changes to fitting window width and the smoothing baseline as well as (ii) the strength of the modulation relative to the local noise. Once the modulations are constructed, we measure periods as intervals between consecutive extrema (peak-to-peak and trough-to-trough), resulting in ten periods. The uncertainty on each period is constructed by combining the uncertainties of each component endpoint in quadrature. We note that periods which share an endpoint are anticorrelated and therefore any fits performed using the periods employ the full covariance matrix. 

\subsection{Red noise analysis}
\label{sec:rednoise_methods}
% Description of test to determine probability that SN 2024afav modulations were generated by red noiseincluding test at 2 levels: SLSN sample to inform red noise distribution AND 24afav-only (most conservative)
% FIGURE: whatever makes the most sense for reporting the red noise result (e.g., sample realizations, probability distributions, etc.)

A counter to the claim that the modulations represent a coherent physical process is that stochastic variability with a steep (red) power spectrum can produce several cycles of apparently periodic behavior (see e.g., the problem of identifying periodic signals in red-noise-contaminated active galactic nuclei light curves, \cite{Krishnan2021}). The modulations in other SLSNe-I are potentially driven by CSM interaction, so in principle, if SN 2024afav is also driven by CSM, its light curve modulations should be similar to others in its object class. We evaluate this claim for SN 2024afav by simulating red noise light curves at the photometric cadence and uncertainties of SN 2024afav and characterizing how often a chirp-antichirp appears in the randomly generated light curves. We calibrate our red-noise model to six objects from the SLSN-I population (SN 2007bi, SN 2017gci, SN 2018gkz, SN 2018ibb, SN 2020qlb, and SN 2020wnt; data taken from \cite{Gomez2024}) which have modulations as well as well-sampled $r$-band photometry beyond $150$ d post-peak. We subtracted from each the same smooth baseline used for SN 2024afav and fitted a power-law spectrum $P(f)\propto f^{-\alpha_{\rm ps}}$ to the residuals, and adopted the median amplitude and slope ($\alpha_{\rm ps}=2.85$). For each model, we generated red-noise realizations with the method of \cite{Timmer1995}. We added each realization to the smoothed baseline of SN 2024afav at the observed $r$-band epochs and incorporated white noise whose per-epoch magnitude was drawn from SN 2024afav's photometric errors. Each simulated light curve was then passed through the identical pipeline used on SN 2024afav (nightly binning, baseline subtraction per band, automated extremum extraction). We retained only realizations whose number of extracted extrema lay within $\pm2$ of SN 2024afav, up to a total of $\sim5\times10^4$ realizations of the model. 

Each light curve (real and simulated) was then characterized by the following two quantities: (i) modulation power, measured as the excess rms of the residual over the span of the measured extrema, after subtracting the photometric error variance; and (ii) parabola ``contrast'', a measure of how cleanly the period sequence traces a chirp followed by an antichirp. To construct the parabola contrast, we constructed the ``period'' evolution from the extracted extrema and fitted a parabola to it with Cauchy loss. The contrast is then the rise of the shallower arm of the parabola above the vertex, divided by the scatter of the points about the fit. We rejected realizations which produced parabolas with non-positive or insignificant curvature, or a vertex that lay outside the period sequence, with at least two points pre- and post-minimum. Under this prescription, SN 2024afav has a modulation power of 0.22 mag and a parabola contrast of 13.0 (corresponding to a 14.7 d rise in period on the antichirp arm against a scatter of 1.1 d). For each quantity, we computed the fraction of realizations at least as extreme as SN 2024afav and combined them through their product, and calculated a combined $p$-value as the fraction of noise realizations whose product $p_{\rm power}p_{\rm contrast}$ is at least as small as SN 2024afav. Against the population-calibrated noise, red noise matches or exceeds SN 2024afav's modulation power ($0.22$ mag) within just 74 of $5\times10^4$ realizations ($p_{\rm power}=1.5\times10^{-3}$) and its parabola contrast ($13.0$) in $1{,}280$ ($p_{\rm contrast}=0.026)$. Just 31 realizations are as or more extreme than SN 2024afav in the combined test, a combined $p$-value of $6.4\times10^{-4}$ ($\sim3.2\sigma)$. SN 2024afav is therefore an extreme outlier even among red-noise light curves calibrated on other bumpy SLSNe-I. 

To perform a more conservative test, we run the same analysis as above but using a red-noise power spectrum calibrated only to the modulations of SN 2024afav. By construction, this method negates the rarity based on modulation power (as amplitude is calibrated to the mean amplitude of SN 2024afav, the only object in the sample). Despite this, red noise calibrated specifically to SN 2024afav's power spectrum yields extrema with comparable parabola contrast $\ll5\%$ of the time. 

\begin{figure}
\centering
\includegraphics[width=\textwidth]{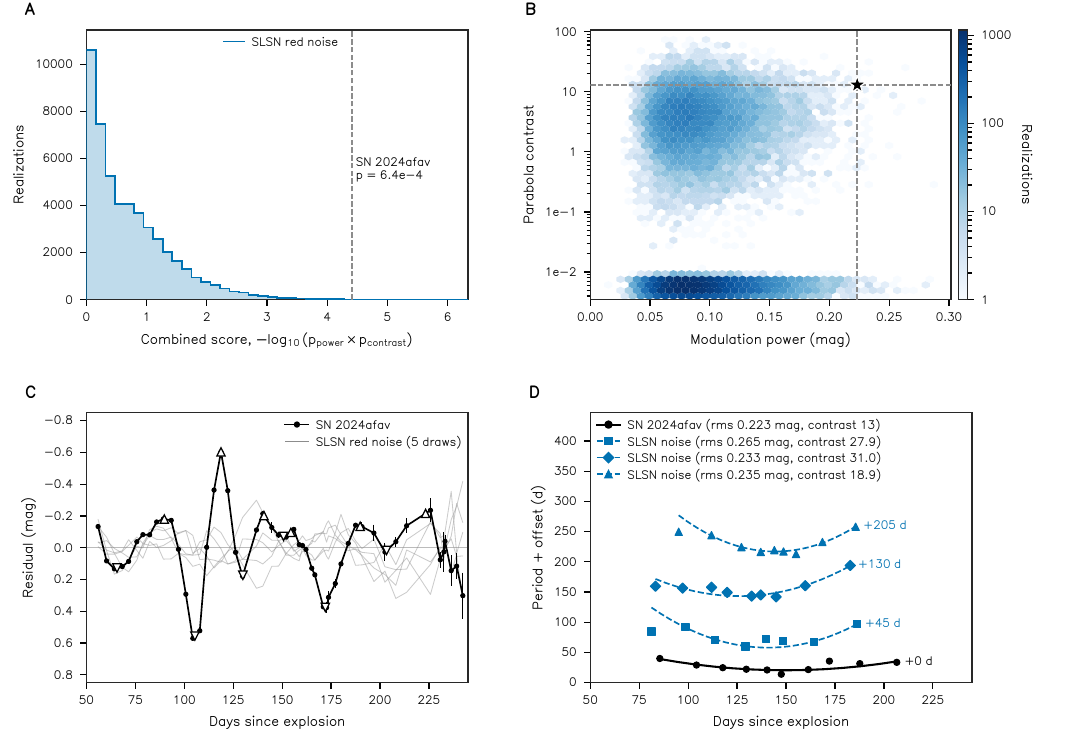}
\caption{\textbf{Are the SN 2024afav modulations likely to be red noise?} A common explanation for modulations in SLSNe-I is interaction with randomly-distributed shells of CSM. We test this explanation against the modulations observed in SN 2024afav, which have two unique features relative to other SLSNe. First, they are unusually high-amplitude, and second, they produce a chirp followed by an antichirp. We constructed a synthetic red noise power spectrum by modeling the post-peak modulations of six SLSNe-I as red noise and generating $\sim5\times10^4$ realizations similar to SN 2024afav's photometric uncertainty and cadence (examples shown in panel C). We then measured the rms amplitude and extracted extrema timings in an identical manner to our procedure for SN 2024afav. We find that that SN 2024afav is an extreme outlier even in a red noise distribution tuned to modulations in SLSNe-I. The two-dimensional distribution of realizations in parabola contrast and modulation power is shown in panel B; the distribution of a joint score of rarity (constructed based on the $p$-value of each realization relative to the overall marginal distribution in modulation power or parabola contrast, then combining those scores via a log product) is shown in panel A. Fewer than 100 realizations were more extreme than SN 2024afav; a sample of these few are shown in panel D.}
\label{fig:rednoise_fig}
\end{figure}

\section{Supplementary Text}
\label{sec:supptext}

\subsection{Validation of chirp-antichirp behavior}
\label{sec:validation_chirpantichirp_supp}
% Multiple validations/cross-checks for the quasiperiodic behavior

To validate the presence of the quasiperiodicity (particularly, the chirp-antichirp) further, we investigate the period evolution without reference to individual extrema, using a time-frequency decomposition of the residual. We first scale the residual by a smooth envelope to ensure every cycle contributes equally, then computed the weighted wavelet Z-transform, suitable for unevenly-sampled data, following the method of \cite{Foster1996}, using 240 logarithmically-spaced frequencies between periods of 9 and 48 d. The resulting map is then refined via the method of synchrosqueezing \cite{daubechies1996nonlinear,daubechies2011synchrosqueezed}, which reassigns each cell's power to the instantaneous frequency implied by how the fitted phase advances with time. Finally, each time column is normalized to its maximum, to assist with interpretability and visualization. 

The resulting map shows two clear ridges which evolve with time. The dominant ridge rises from $\sim0.025 \textrm{ d}^{-1}$ near $t\sim90$ d to $\sim0.07 \textrm{ d}^{-1}$ near $t\sim163$ d, then turns over and sharply descends. The ridge is identified (up to $\sim7\%$ in frequency) with the locations of the extrema, confirming that the visible extrema track the dominant mode in the power spectrum. The time-frequency decomposition recovers the chirp and subsequent reversal independently of how the extrema were selected or extracted. We also note the presence of a lower ridge at periods of $\sim33-39$ d, which traces a slower variation observable in the modulations, namely the alternating deep and shallow troughs. 

\subsection{Selection of discrete menu for the universal relation}
\label{sec:discrete_supp}
% explanation of how the choices for the discrete menu were made

\subsection*{Disk prescriptions}

Both prescriptions share the assumptions of the universal relation, $L_w/r^2 \propto P_{\rm disk}(r)$, $\dot M \propto t^{-\chi}$, $L_w = -I\Omega\dot\Omega$, and precession period $P \propto J^{-q} r^{K}$. The disk enters
only through $P_{\rm disk} \propto \dot M^{u} r^{-w}$, with $s \equiv 1/(w-2)$, so that the full relation reads
\begin{equation}
C = K s\,\chi u - \left[K s + \frac{q}{n+1}\right] D .
\label{eq:UR_full}
\end{equation}

\paragraph{Thin disk.}
We adopt the steady, geometrically-thin $\alpha$-disk relations \cite{FKR}: $\rho = \Sigma/H$, $H = c_s R^{3/2}(GM)^{-1/2}$, $T^4 = \frac{9\tau\, GM\dot M R^{-3}}{32\pi\sigma}\,$, $c_s^2 = \gamma P/\rho$, $\tau = \Sigma\kappa$, $\nu\Sigma = \dot M/3\pi$, and $\nu = \alpha c_s H$. Here $\dot M$ is independent of radius (no outflow), and we work at $R \gg R_*$, where the inner-boundary factor $1-(R_*/R)^{1/2} \to 1$. We close the system with a power-law equation of state, $P_{\rm disk} \propto \rho^{A} T^{B}$, and opacity, $\kappa \propto \rho^{C} T^{D}$. Gas (radiation) pressure has $(A,B) = (1,1)$ [$(0,4)$], and Kramers' (electron-scattering) opacity has $(C,D) = (1,-7/2)$ [$(0,0)$] \cite{1996ima..book.....C}. For any $(A,B,C,D)$ the system can be solved algebraically for $T$, $\rho$, $\Sigma$, and $v_r = -\dot M/(2\pi R\Sigma)$ as pure power laws in $\alpha$, $\dot M$, $M$, and $R$. Each pressure component, $P_{\rm gas} = (k_B/\mu m_p)\rho T$, $P_{\rm rad} = (4\sigma/3c)T^4$, and $P_{\rm ram} = \tfrac{1}{2}\rho v_r^2$, then scales as $P_{\rm disk} \propto \dot M^{u} R^{-w}$. The quantities $\alpha$, $M$, $\mu$, and all numerical coefficients appear only in the normalization, which cancels in the universal relation. Because $\dot M$ is constant with radius, each regime and component fixes a single $(u,w)$ with no continuous parameter
(Table~\ref{tab:thin_menu}).

\begin{table}
\centering
\caption{\textbf{Thin-disk closure menu.} Here, we organize exponents of $P_{\rm disk} \propto \dot M^{u} R^{-w}$ for each disk regime and pressure component, with $s = 1/(w-2)$. Entries with $s<0$ cannot produce a chirp followed by an antichirp.}
\begin{tabular}{llccc}
\hline
Regime & Component & $u$ & $w$ & $s$ \\
\hline
Gas + Kramers'  & $P_{\rm gas}$ & $17/20$ & $21/8$  & $8/5$   \\
                & $P_{\rm rad}$ & $6/5$   & $3$     & $1$     \\
                & $P_{\rm ram}$ & $23/20$ & $19/8$  & $8/3$   \\
Gas + e.s.      & $P_{\rm gas}$ & $4/5$   & $51/20$ & $20/11$ \\
                & $P_{\rm rad}$ & $8/5$   & $18/5$  & $5/8$   \\
                & $P_{\rm ram}$ & $6/5$   & $49/20$ & $20/9$  \\
Rad + Kramers'  & $P_{\rm gas}$ & $2/25$  & $9/5$  & $-5$  \\
                & $P_{\rm rad}$ & $16/25$ & $12/5$  & $5/2$   \\
                & $P_{\rm ram}$ & $34/25$ & $13/5$  & $5/3$   \\
Rad + e.s.      & $P_{\rm gas}$ & $-2$    & $-9/8$  & $-8/25$ \\
                & $P_{\rm rad}$ & $0$     & $3/2$   & $-2$    \\
                & $P_{\rm ram}$ & $2$     & $7/2$   & $2/3$   \\
\hline
\end{tabular}
\label{tab:thin_menu}
\end{table}

\paragraph{Thick disk.}
We assume a self-similar advective flow, which implies constant $H/R \sim c_s/v_K$ \cite{Narayan1994}. That solution was derived for optically thin, low-$\dot M$ flows, but gravity is the only scale in both cases, so the specific scalings for mass loss apply. The scalings with mass loss are therefore those of \cite{Yuan2014} (eqs.~8--11). Mass is lost via an outflow that depends on radius, $\dot M = \dot M_0 (r/r_{\rm out})^{p}$ with $p \in [0,1)$ \cite{Blandford1999}. We propagate this into the surface density, $\Sigma \propto \dot M/(r|v_r|) \propto \dot M(r)\, r^{-1/2}$, and the density, $\rho \propto \Sigma/H \propto \dot M(r)\, r^{-3/2}$. The total disk pressure is then $P_{\rm disk} \propto \rho c_s^2 \propto \dot M(r)\, r^{-5/2} \propto \dot M_0\, r_{\rm out}^{-p}\, r^{p-5/2}$.

The final assumption we require is the radius $r_{\rm out}$ at which the outflow begins. In a super-Eddington disk the flow becomes radiatively inefficient inside the photon-trapping (spherization) radius, so $r_{\rm out} = r_{\rm tr} \propto \dot M_0$ \cite{2007MNRAS.377.1187P}. With this choice, the pressure at fixed radius scales as $\dot M_0^{u}$ with $u = 1-p$. Similarly, $w = 5/2 - p$, which implies $s = 1/(1/2 - p)$; a constant $r_{\rm out}$ instead gives $u = 1$. This introduces a continuous parameter $p$. Simulations of advective flows with outflows find $p \approx 0.4$--$0.7$ \cite{Yuan2012}. Because the chirp + antichirp configuration is only supported by $s > 0$, we must have $p < 1/2$.

Substituting into Eq.~\eqref{eq:UR_full},
\begin{equation}
C = \frac{K}{\tfrac{1}{2}-p}\,\chi(1-p)
  - \left[\frac{K}{\tfrac{1}{2}-p} + \frac{q}{n+1}\right] D .
\end{equation}
The choice of $p$ rotates the relation around a single point. For given $K$, $\chi$, $q$, and $n$, every $p$ yields a line through
\begin{equation}
(D_{\rm pivot},\, C_{\rm pivot}) =
\left(\frac{\chi}{2},\; \chi\left[K - \frac{q}{2(n+1)}\right]\right).
\end{equation}
The linearity of the relation, combined with this fixed point, lets us keep its discreteness even though $p$ is a continuous parameter. This holds whether the precession radius is set by the inner edge or by the entire zone precessing rigidly.

\begin{figure}
\centering
\includegraphics[width=\textwidth]{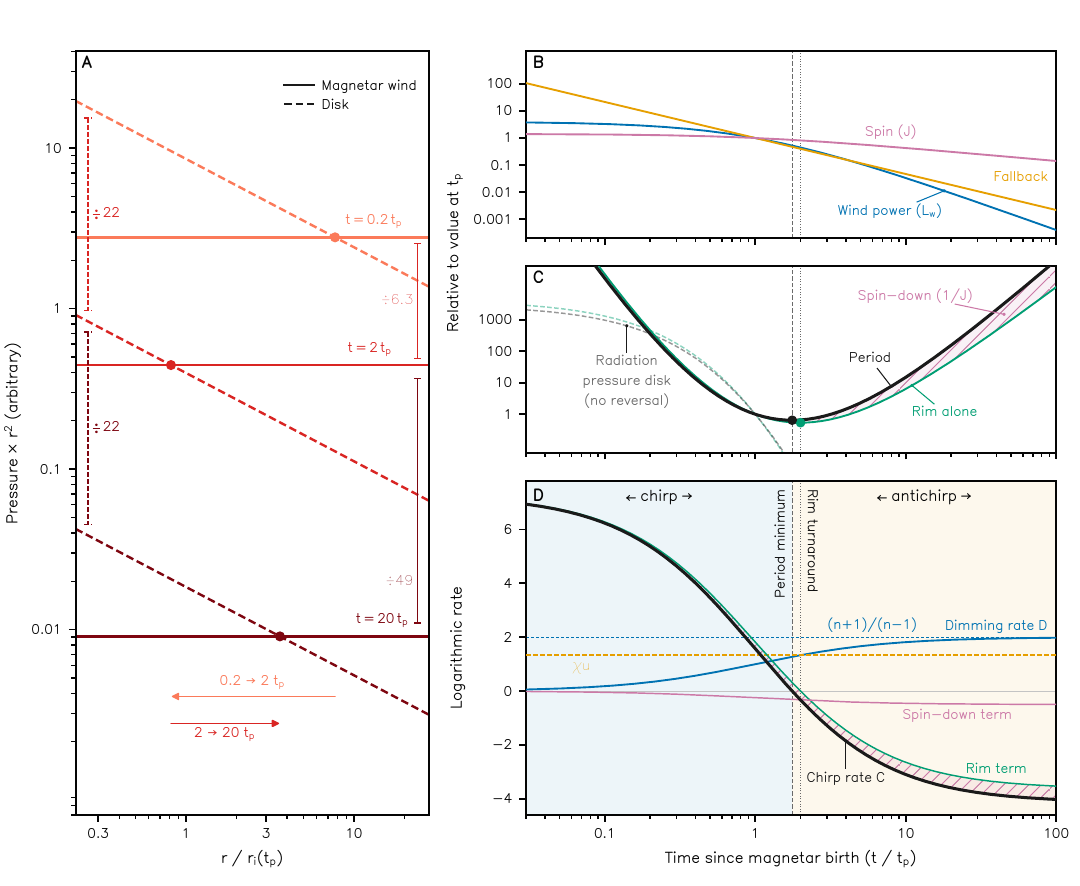}
\caption{\textbf{Physical evolution of the system.} We visualize different aspects of the proposed model dynamics to reveal how a chirp and subsequent antichirp arise. The radius of the disk informs the precession frequency, and is set by the balance point between the magnetar wind and the pressures of the disk (Panel A). At early times, the disk pressure decreases at a higher rate than the magnetar output, leading to the radius shrinking towards the magnetar (shrinking radius $\to$ increasing precession frequency, i.e., chirp). After a characteristic spin-down time of the magnetar ($t\gtrsim t_p$), the magnetar engine output decreases rapidly and outpaces the decline of the disk pressure. This interplay (Panel B) pushes the disk out to higher radii, resulting in the antichirp. The universal relation highlights that this turnaround should occur when the dimming rate $D$ crosses $\approx\chi u$, corresponding to the engine fading faster than the pressures within the disk. The location of the turnaround is only mildly affected by the term encoding the changing angular momentum of the magnetar associated with the Lense-Thirring precession mechanism; the behavior is primarily driven by the interaction between the disk pressure and the magnetar wind (curves labeled ``rim''). An intriguing aspect is that the appearance of the chirp-antichirp occurs only for some disk configurations and not others. For example, the disk visualized here is gas-pressure dominated; but a radiation-pressure-dominated disk (Panel C, dashed lines) features only a chirp, but no reversal. }
\label{fig:reversal_physics}
\end{figure}

\begin{figure}
\centering
\includegraphics[width=0.5\textwidth]{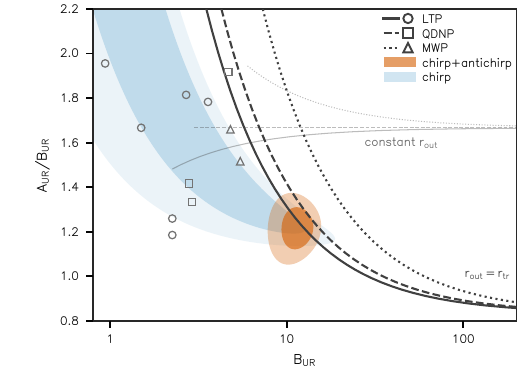}
\caption{\textbf{Universal relation constraints from SN 2024afav.} We visualize the constraint provided by SN 2024afav in an alternative manner to the previous $(D,C)$ plot. Here, we consider the parameter space of allowed and reported slopes ($B_{\rm UR}$) and slope-intercept ratios ($A_{UR}/B_{UR}$). The thin disk prescription results in one $(D,C)$ line per combination of system parameters, which results in single points in this view. The SN 2024afav chirp (blue contours) is consistent with the thin disk prescription comb, which is why the published model reported consistency. However, the presence of the antichirp firmly disfavors any thin-disk configuration. When we instead examine the thick disk prescription (whose continuous $p$ value from $\dot M\propto r^p$ results in each configuration tracing a line in this view instead of a single point), we find that SN 2024afav is consistent. A further benefit of this transformation of the constraints is that certain model choices have a significant impact; for example, fixing the outer radius of the disk makes the thick disk prescription completely inconsistent with SN 2024afav when compared to setting the outer disk radius equal to the photon trapping radius.}
\label{fig:ab_vs_b_supp}
\end{figure}

\section{Supplementary Figures and Tables}
\label{sec:suppfigs}

\begin{table}[h]
\caption{\textbf{Extrema of the modulations of SN\,2024afav.} Epochs are observer-frame MJD; $t$ is observer-frame days since explosion (MJD 60646.08), as in Fig.~1. Each epoch is the vertex of a local quadratic fit to the residual, seeded at a visual identification. $\sigma_{\rm fit}$ combines the statistical, window and baseline terms; $\sigma_{\rm amp}=(P/2\pi)(\sigma_{\rm noise}/a)$ is the amplitude term; $\sigma_t$ is their quadrature sum. We begin counting from the second extremum visible in the data (labeled E02 below), as the first extremum is coincident with the peak of the supernova.}
\label{tab:extrema}
\begin{ruledtabular}
\begin{tabular}{llrrrrr}
ID & Type & MJD & $t$ (d) & $\sigma_{\rm fit}$ (d) & $\sigma_{\rm amp}$ (d) & $\sigma_t$ (d) \\
\hline
E02 & trough & 60725.59 & 79.51  & 7.34 & 1.51 & 7.49 \\
E03 & peak   & 60747.75 & 101.67 & 1.57 & 0.59 & 1.68 \\
E04 & trough & 60765.22 & 119.14 & 1.13 & 0.31 & 1.17 \\
E05 & peak   & 60778.82 & 132.74 & 1.10 & 0.28 & 1.14 \\
E06 & trough & 60791.26 & 145.18 & 1.12 & 0.37 & 1.18 \\
E07 & peak   & 60803.39 & 157.32 & 1.37 & 0.78 & 1.58 \\
E08 & trough & 60811.48 & 165.40 & 2.64 & 1.94 & 3.28 \\
E09 & peak & 60820.75 & 174.67 & 4.59 & 1.21 & 4.74 \\
E10 & trough & 60835.94 & 189.86 & 1.22 & 0.86 & 1.49 \\
E11 & peak   & 60853.37 & 207.29 & 1.37 & 0.86 & 1.61 \\
E12 & trough & 60874.08 & 228.00 & 1.61 & 0.94 & 1.87 \\
E13 & peak   & 60888.88 & 242.80 & 2.24 & 3.90 & 4.50 \\
\end{tabular}
\end{ruledtabular}
\end{table}

\begin{table}[h]
\caption{\textbf{Modulation periods.} P01--P10 are intervals between consecutive extrema of the same type (Table~\ref{tab:extrema}); $\sigma_P$ combines the two endpoint uncertainties in quadrature, and periods two apart share an endpoint. Observer frame: midpoint $t_{\rm obs}$ in days since explosion (MJD 60646.08), as in Fig.~2C. Rest frame: the convention of Fig.~3, i.e.\ days since explosion divided by $1+z=1.072$, with the midpoint moved earlier by the diffusion shift $\Delta$ (rest-frame days).}
\label{tab:periods}
\begin{ruledtabular}
\begin{tabular}{llrcrrc}
& & \multicolumn{2}{c}{Observer frame} & \multicolumn{3}{c}{Rest frame} \\
\cline{3-4}\cline{5-7}
ID & Endpoints & $t_{\rm obs}$ (d) & $P\pm\sigma_P$ (d) & $\Delta$ (d) & $t_{\rm rest}$ (d) & $P\pm\sigma_P$ (d) \\
\hline
P01 & E02--E04 & 99.33  & $39.63\pm7.58$ & 7.76 & 84.90  & $36.97\pm7.07$ \\
P02 & E03--E05 & 117.20 & $31.07\pm2.03$ & 6.58 & 102.76 & $28.98\pm1.89$ \\
P03 & E04--E06 & 132.16 & $26.04\pm1.66$ & 5.83 & 117.45 & $24.29\pm1.55$ \\
P04 & E05--E07 & 145.03 & $24.58\pm1.95$ & 5.32 & 129.97 & $22.93\pm1.82$ \\
P05 & E06--E08 & 155.29 & $20.23\pm3.48$ & 4.96 & 139.90 & $18.87\pm3.25$ \\
P06 & E07--E09 & 166.00 & $17.36\pm5.00$ & 4.64 & 150.20 & $16.19\pm4.66$ \\
P07 & E08--E10 & 177.63 & $24.46\pm3.60$ & 4.34 & 161.36 & $22.81\pm3.36$ \\
P08 & E09--E11 & 190.98 & $32.62\pm5.01$ & 4.04 & 174.12 & $30.43\pm4.67$ \\
P09 & E10--E12 & 208.93 & $38.15\pm2.39$ & 3.69 & 191.21 & $35.58\pm2.23$ \\
P10 & E11--E13 & 225.05 & $35.51\pm4.78$ & 3.43 & 206.51 & $33.13\pm4.46$ \\
\end{tabular}
\end{ruledtabular}
\end{table}

\end{document}